\documentclass[aps,showpacs,twocolumn]{revtex4}
\usepackage{epsfig}
\usepackage{amsmath}
\usepackage{graphicx}
\begin{document}

\title{Analysis of $\Omega(2012)$ as a molecule in the chiral quark model}
\author{Xiaohuang Hu$^1$, Jialun Ping$^+$}

\affiliation{Department of Physics and Jiangsu Key Laboratory for Numerical
Simulation of Large Scale Complex Systems, Nanjing Normal University, Nanjing 210023, P. R. China}

\begin{abstract}
Inspired by the updated information on $\Omega(2012)$ by the Belle Collaboration, we conduct a study of all possible $S$-wave pentaquark systems with
quark contents $sssq\bar{q},q=u,d$ in a chiral quark model with the help of Gaussian expansion method. Channel coupling is also considered.
The real-scaling method (stabilization  method) is employed to identify and check the bound states and the genuine resonances. In addition,
the decay widths of all resonances are given. The results show that $\Omega(2012)$ can be interpreted as a $\Xi^*K$ molecular state with quantum number
of $IJ^P=0(\frac{3}{2})^-$. Other resonances are obtained: $\Xi^* K^*$ with $IJ^P=0(\frac{1}{2})^-$ and$0(\frac{3}{2})^-$, $\Omega\pi$ with
$IJ^P=1(\frac{3}{2})^-$. These pentaquark states is expected to be further verified in future experiments.

\end{abstract}


\maketitle

\section{Introduction}

In 2018, an exited state of $\Omega$ with a significance of 8.6$\sigma$ was reported in the $K^-\Xi^0$ and $K^0_S\Xi^-$ invariant mass distributions by
the Belle Collaboration~\cite{1}. The measured mass and width of the state are $2012.4 \pm 0.7 \pm 0.6$ MeV and $6.4^{+2.5}_{-2.0} \pm1.6$ MeV, respectively.
Although the ground state $\Omega$(1672) was predicted in the quark model with the SU(3)-flavor symmetry by Gell-Mann~\cite{2} and Ne'eman~\cite{3} and
discovered experimentally about half-century ago, only three $\Omega$ exited states were listed in the Particle Data Group before the observation by
the Belle Collaboration and there was little knowledge of the nature of them. Thus, the discovery of $\Omega$(2012) was mostly welcomed and triggered many theoretical investigations on the nature of the exited states of $\Omega$.

Before the observation of the state $\Omega$(2012), a series of theoretical models and approaches, such as the quark model~\cite{4,5}, Skyrme model~\cite{6},
lattice QCD~\cite{7}, and large $N_c$~\cite{8}, predicted that the masses of the first excited states of the $\Omega$(1672) are around 2.0 GeV, which are
consistent with the result of the Belle Collaboration. Thus, it is reasonable to explain the $\Omega$(2012) as a good candidate for the first orbital excited
state of $\Omega$(1672). After the observation of $\Omega$(2012), the subsequent theoretical researches have further explored the $qqq$ picture of $\Omega$(2012)
in the chiral quark model~\cite{9}, QCD sum rule~\cite{10} and $^3P^0$ model~\cite{11}.

Nevertheless, the possibility of the pentaquark picture of $\Omega$(2012) cannot be excluded, considering its mass is very close to the threshold of the
state $\Xi^* \bar{K}$. Actually, various theoretical approaches~\cite{12,13,14,15} have investigated the $\Omega$ excited states in pentaquark picture
before the Belle Collaborations's report. The observation of the $\Omega$(2012) also triggered a lot of theoretical work on this issue.
In Ref.~\cite{16}, the flavor SU(3) analysis was performed and found that the $\Omega$(2012) is likely to be a $\Xi^* \bar{K}$ molecular state with $J^p=\frac{3}{2}^-$. Several work also interpreted the $\Omega$(2012) as a possible $\Xi^* \bar{K}$ molecule state and further predicted a large decay width
for the $\Omega(2012)\rightarrow\Xi^* \bar{K}\rightarrow\Xi\pi\bar{K}$~\cite{17,18,19,20}. In Ref.~\cite{21}, the $\Omega$(2012) was a dynamically generated
state from the coupled channel interactions of $\Xi^* \bar{K}$ and $\Omega\eta$ in $S$ wave. Inspired by these theoretical results, a follow-up experiment was
reported by the Belle Collaboration to measure three body decay of the $\Omega$(2012) to $\Xi\pi\bar{K}$~\cite{22}. The result showed that there is no significant
$\Omega$(2012) signals in the $\Xi\pi\bar{K}$ channel, which was in tension with molecular interpretation of the $\Omega$(2012).
Based on the measurements, Refs.~\cite{23,24} revisited the $\Omega$(2012) by the coupled channel unitary approach in the molecular perspective, taking account
of the interaction of the $\Xi^* \bar{K}$, $\Omega\eta$, and $\Xi K$(D-wave) channels and indicated that the experimental properties of $\Omega$(2012)~\cite{22}
can be easily accommodated. In the hadronic molecular approach, the state $\Omega$(2012) can be interpreted as the $P$-wave $\Xi^*K$ molecule state with
$IJ^P=0(\frac{1}{2})^+$ or $\frac{3}{2}^+$~\cite{25}, while in Ref.~\cite{26}, the $\Omega$(2012) was considered to contain the mixed $\Xi^*\bar{K}$ and
$\Omega\eta$ hadronic components. In Ref.~\cite{27}, a nonrelativistic constituent quark potential model was used and $\Omega$(2012) can be interpreted
as a $P$-wave state with $J^P=\frac{3}{2}^-$ in $qqq$ picture. In Ref.~\cite{28}, the $\Omega$(2012) was study in the nonleptonic weak decays of
$\Omega_c^0\rightarrow\Xi^*\bar{K}\pi^+(\Omega\eta)\rightarrow(\Xi\bar{K})^-\pi^+$ and $(\Xi\bar{K}\pi)^-\pi^+$ via final-state interactions of the
$\Xi^*\bar{K}$ and $\Omega\eta$ pairs. However, the Belle Collaboration has revisited the measurement of $\Omega(2012)^-\rightarrow\Xi(1530)^0
K^-\rightarrow\Xi^-\pi^+K^-$ and update the measurements of $\Omega(2012)^-\rightarrow\Xi^0 K^-$ and $\Omega(2012)^-\rightarrow\Xi^- K_S^0$ with improved
selection criteria very recently~\cite{29}. The results show that the ratio of the branching fraction for the resonant three-body decay to that for
the two-body decay to $\Xi\bar{K}$ is $0.97\pm0.24\pm0.07$, consistent with the molecular picture of $\Omega(2012)$. The conclusion is in contrast to
the previous study~\cite{22} and suggests that $\Omega(2012)$ can be interpreted as a molecular state.

The state $\Omega$(2012) was also investigated in the framework of chiral quark model (ChQM) and quark delocalization color screening model (QDCSM)
before in our group~\cite{30}. However, there is a mass inversion of $\Xi^*\bar{K}$ channel and $\Omega\eta$ channel due to the single Gaussian approximation
in constructing the wave functions of hadrons. In the Rayleigh-Ritz variational method, the basis expansion of the trial wave function is significant.
In this work, Gaussian expansion method (GEM)~\cite{31} in which each relative motion in the system is expanded in terms of Gaussian basis functions
whose sizes are taken in geometric progression, is adopted. The GEM has proven to be an accurate and universal few-body calculation method~\cite{32,33,34,35}
and we hope to solve the previous problem of mass inversion by this method and to describe the hadron spectrum better. The constituent chiral quark model
will still be employed to investigate all possible $S$-wave pentaquark systems for $\Omega$(2012), considering the channel coupling interactions.
After considering all possible configurations of color, spin and flavor degrees of freedom, we can identify the structures of these systems. Finally,
with the help of ``real scaling method"~\cite{36,37,38}, we can confirm the bound states and the genuine resonances and obtain decay widths of resonances.

The paper is organized as follow. After introduction, details of ChQM and GEM are introduced in Section II. In Section III, we present the numerical results
and a method of finding and calculating the decay width of the genuine resonance state (``real scaling method").
 Finally, We give a brief summary of this work in the last section.

\section{CHIRAL QUARK model and wave functions}
When it comes to multiquark candidates observed by experiments, QCD-inspired quark model approach is still one of the most common tool for describing them.
The chiral quark model has become one of the most effective approaches to describe hadron spectra, hadron-hadron interactions and multiquark states~\cite{39}.
The general form of five-body Hamiltonian in the model is given as follows:
\begin{eqnarray}
H & =& \sum_{i=1}^{5}\left( m_i+\frac{p^2_i}{2m_i}\right)-T_{CM}  \nonumber  \\
  & +& \sum_{j>i=1}^{5}\left[ V_{CON}({{\bf r}_{ij}})+V_{OGE}({{\bf r}_{ij}})
      +V_{GBE}({{\bf r}_{ij}})\right] ,
\end{eqnarray}
where $m_i$ is the constituent mass of quark (antiquark), $\bf{p}_i$ is momentum of quark and $T_{cm}$ is the kinetic energy of the center-of mass motion.
Due to the fact that a nearly massless current light quark acquires a dynamical, momentum dependent mass (so-called constituent quark mass) for its interaction
with the gluon medium, ChQM contains color confinement potential, one-gluon exchange potential (OGE) and Goldstone boson exchange potentials (GBE).
These three potentials reveal the most relevant features of QCD at low energy regime: confinement, asymptotic freedom and chiral symmetry spontaneous breaking.

In this work, we are interested in the low-lying pentaquark systems of $S$-wave composed of $sssq\bar{q},q=u,d$. Hence, only the central part of the
quark-quark interactions are considered.
For color confinement potential, the screened form is used,
\begin{equation}
\setlength{\abovedisplayskip}{5pt}
\setlength{\belowdisplayskip}{5pt}
V_{CON}({{\bf r}_{ij}})  =  \boldsymbol{\lambda}_i^c\cdot \boldsymbol{\lambda}_j^c
   \left[-a_c (1-e^{-\mu_cr_{ij}})+\Delta \right] ,
\end{equation}
where $a_c$, $\mu_c$ and $\Delta$ are model parameters, and $\boldsymbol{\lambda}^c$ represent the SU(3) color Gell-Mann matrices.

One-gluon exchange potential contains so-called coulomb and color-magnetism interactions, which arise from QCD perturbation effects,
\begin{equation}
V_{OGE}({{\bf r}_{ij}})  =  \frac{1}{4}\alpha_s \boldsymbol{\lambda}_i^c \cdot \boldsymbol{\lambda}_j^c
   \left[ \frac{\boldsymbol{\sigma}_i\cdot \boldsymbol{\sigma}_j}{r_{ij}}-\frac{1}{6m_im_j}
   \frac{e^{-r_{ij}/r_0(\mu)}}{r_{ij}r^2_0(\mu)}\right] ,
\end{equation}
$\mu$ is the reduced mass between two interacting quarks; $\boldsymbol{\sigma}$ represent the SU(2) Pauli matrices; $r_0(\mu)=\hat{r}_0/\mu$;
$\alpha_s$ denotes the effective scale-dependent strong running coupling constant of one-gluon exchange,
\begin{equation}
\alpha_{s} =\frac{\alpha_{0}}{\ln(\frac{\mu^2+\mu_{0}^2}{\Lambda_{0}^2})}.
\end{equation}

Due to chiral symmetry spontaneous breaking, Goldstone boson exchange potentials appear between light quarks ($u$,$d$ and $s$),
\begin{eqnarray}
V_{GBE}({{\bf r}_{ij}})&=&V_{\pi}({{\bf r}_{ij}})+V_{K}({{\bf r}_{ij}})+V_{\eta}({{\bf r}_{ij}})+V_{sc}({{\bf r}_{ij}}), \nonumber \\
V_{\pi}({{\bf r}_{ij}})&=&\frac{g^2_{ch}}{4\pi}\frac{m^2_\pi}{12m_im_j}
	\frac{\Lambda^2_\pi m_\pi}{\Lambda^2_\pi-m^2_\pi}[ Y(m_{\pi}r_{ij}) \nonumber \\
&-& \frac{\Lambda^3_\pi}{m^3_\pi}Y (\Lambda_{\pi}r_{ij}) ]
	(\boldsymbol{\sigma}_i \cdot \boldsymbol{\sigma}_j)\sum_{a=1}^{3}
	\lambda_i^a \lambda_j^a,   \nonumber  \\
V_{K}({{\bf r}_{ij}})&=&\frac{g^2_{ch}}{4\pi}\frac{m^2_K}{12m_im_j}
	\frac{\Lambda^2_K m_K}{\Lambda^2_K-m^2_K}[ Y(m_{K}r_{ij}) \nonumber \\
&-&\frac{\Lambda^3_K}{m^3_K}Y (\Lambda_{K}r_{ij}) ]
	(\boldsymbol{\sigma}_i \cdot \boldsymbol{\sigma}_j)\sum_{a=4}^{7}
	\lambda_i^a \lambda_j^a,    \\
V_{\eta}({{\bf r}_{ij}})&=&\frac{g^2_{ch}}{4\pi}\frac{m^2_\eta}{12m_im_j}
	\frac{\Lambda^2_\eta m_\eta}{\Lambda^2_\eta-m^2_\eta} \nonumber \\
& & [ Y(m_{\eta}r_{ij})-\frac{\Lambda^3_\eta}{m^3_\eta}Y (\Lambda_{\eta}r_{ij}) ]
    (\boldsymbol{\sigma}_i \cdot \boldsymbol{\sigma}_j)  \nonumber \\
& &	[\cos\theta_{P}(\lambda_i^8 \lambda_j^8)-\sin\theta_{P}(\lambda_i^0 \lambda_j^0)],  \nonumber
\end{eqnarray}
$Y(x)$ is the standard Yukawa functions; $\lambda^a$ are the $SU(3)$ flavor Gell-Mann matrices; $\Lambda$ are the cut-offs and
$m_{\chi},\chi=\pi,K,\eta$ are the masses of Goldstone bosons; $g^2_{ch}$ is the chiral field coupling constant,
which is determined from the $NN\pi$ coupling constant through
\begin{equation}
\frac{g^2_{ch}}{4\pi}=\frac{9}{25}\frac{g^2_{\pi NN}}{4\pi}\frac{m^2_{u,d}}{m^2_N}.
\end{equation}
Besides, the scalar nonet (the extension of chiral partner $\sigma$ meson) exchange $V_{sc}$ is also used in this work, which introduces other higher
multi-pion terms that are simulated through the full nonet of scalar mesons exchange between two constituent quarks~\cite{40}.
\begin{eqnarray}
V_{sc}({{\bf r}_{ij}})&=&V_{a_0}({{\bf r}_{ij}})\sum_{a=1}^{3}
	\lambda_i^a \lambda_j^a+V_{\kappa}({{\bf r}_{ij}})\sum_{a=4}^{7}
	\lambda_i^a \lambda_j^a \nonumber \\
&+& V_{f_0}({{\bf r}_{ij}})
	\lambda_i^8 \lambda_j^8+V_{\sigma}({{\bf r}_{ij}})\lambda_i^0 \lambda_j^0, \nonumber  \\
V_{s}({{\bf r}_{ij}}) & =& -\frac{g^2_{ch}}{4\pi} \frac{\Lambda^2_s m_s}{\Lambda^2_s-m^2_s}
	[ Y(m_{s}r_{ij})-\frac{\Lambda_s}{m_s}Y(\Lambda_{s}r_{ij})], \nonumber  \\
& & ~~~~s=a_0,\kappa,f_0,\sigma
\end{eqnarray}

The model parameters are listed in Table \ref{para}, and the calculated baryon and meson masses are presented in the Table \ref{mass} along with
the experimental values. Because it is difficult to use the same set of parameters to have a good description of baryon and meson spectra simultaneously,
we treat the strong coupling constant of one-gluon exchange with different values for quark-quark and quark-antiquark interacting pairs.
From the calculation, most of the results are close to experimental values except for $\omega$. In the follow-up calculation,
we find that the molecular states consisting of this meson are open channels and these channels have no effect on all stable molecular states
so these errors do not affect our final results.
\begin{table}[h]
\caption{Quark model parameters.\label{para}}
\begin{tabular}{c|cc cc} \hline \hline
Quark masses       &$m_u$=$m_d$ (MeV)   &~~~~313\\
                   &$m_s$ (MeV)  &~~~~555\\  \hline
                   &$\Lambda_\pi$ (fm$^{-1}$)  &~~~~4.20\\
                   &$\Lambda_\eta=\Lambda_K$ (fm$^{-1}$)      &~~~~5.20\\
                   &$m_\pi$ (fm$^{-1}$)  &~~~~0.70\\
Goldstone bosons   &$m_K$ (fm$^{-1}$)  &~~~~2.51\\
                   &$m_\eta$ (fm$^{-1}$)  &~~~~~2.77\\
                   &$g^2_{ch}/(4\pi)$  &~~~~0.54\\
                   &$\theta_P(^\circ)$  &~~~~-15\\  \hline
                   &$a_c$ (MeV)  &~~~~465.3\\
     Confinement       &$\mu_c$ (fm$^{-1})$  &~~~~0.58\\
                    &$\Delta$ (MeV)  &~~~164.52\\   \hline
                   &$m_\sigma$ (fm$^{-1}$)  &~~~~3.42\\
                   &$\Lambda_\sigma$ (fm$^{-1}$)  &~~~~4.20\\
scalar nonet       &$\Lambda_{a_0}=\Lambda_\kappa=\Lambda_{f_0}$ (fm$^{-1}$)  &~~~~5.20\\
                   &$m_{a_0}=m_\kappa=m_{f_0}$ (fm$^{-1}$)  &~~~~4.97\\  \hline
                 &$\hat{r}_0~$(MeV~fm)  &~~~~35.19\\
                    &$\alpha_{uu}$  &~~~~0.467/0.662\\
        OGE           &$\alpha_{us}$  &~~~~0.695/0.573\\
                    &$\alpha_{ss}$  &~~~~0.29/0.335\\   \hline \hline
\end{tabular}
\end{table}

\begin{table}[h]
\caption{The masses of ground-state baryons and mesons (unit: MeV).\label{mass}}
\begin{tabular}{ccccccc}
\hline \hline
 &$\Lambda$~ &$\Sigma$~ &$\Sigma^*$~ &$\Xi$~ &$\Xi^*$~  &$\Omega$~   \\ \hline
CHQM~ &1115~ &1191~ &1392~ &1318~ &1537~ &1672~  \\
Expt~ &1116~ &1189~ &1386~ &1315~ &1530~ &1672~  \\ \hline

&$\pi$~ &$\rho$~ &$\bar{K}$~ &$\bar{K}^*$~ &$\eta'$~ &$\phi$  \\ \hline
CHQM~ &139~ &768~ &498~ &915~ &958~ &1048~ \\
Expt~ &140~ &775~ &498~ &892~ &958~ &1020~ \\
 &$\eta$~ &$\omega$  \\ \hline
CHQM~  &560~ &590~ \\
Expt~  &548~ &782~ \\
\hline \hline
\end{tabular}
\end{table}

In the following, the wave functions for the pentaquark systems are constructed and the eigen-energy is obtained by solving the Schr\"{o}dinger equation.
The wave function of the system consists of four parts: orbital, spin, flavor and color. The wave function of each part is constructed in two steps,
first construct the wave function of three-quark cluster and quark-antiquark cluster, respectively, then coupling two clusters wave functions to form the
complete five-body one.
In this work, the wave functions for $(ssq)(\bar{q}s)$ configuration is written down, the wave functions for other configuration can be obtained
by exchange the indices of particles. The indices of particles $s,s,q,\bar{q},s$ are 1,2,3,4,5. As an example the wave functions for $(sss)(\bar{q}q)$
are obtained by exchange the particle indices $3\leftrightarrow 5$.

The first part is orbital wave function. A five-body system have four relative motions so it is written as follows.
\begin{equation}
\psi^x_{LM_L}=\left[ \left[ \left[
  \psi_{n_1l_1}(\mbox{\boldmath $\rho$})\psi_{n_2l_2}(\mbox{\boldmath $\lambda$})\right]_{l}
  \psi_{n_3l_3}(\mbox{\boldmath $r$}) \right]_{l^{\prime}}
  \psi_{n_4l_4}(\mbox{\boldmath $R$}) \right]_{LM_L},
\end{equation}
where the Jacobi coordinates are defined as follows,
\begin{eqnarray}
{\mbox{\boldmath $\rho$}} & = & {\mbox{\boldmath $x$}}_1-{\mbox{\boldmath $x$}}_2, \nonumber \\
{\mbox{\boldmath $\lambda$}} & = & (\frac{{m_1\mbox{\boldmath $x$}}_1+{m_2\mbox{\boldmath $x$}}_2}{m_1+m_2})-{\mbox{\boldmath $x$}}_3,  \nonumber \\
{\mbox{\boldmath $r$}} & = & {\mbox{\boldmath $x$}}_4-{\mbox{\boldmath $x$}}_5,  \\
{\mbox{\boldmath $R$}} & = & (\frac{{m_1\mbox{\boldmath $x$}}_1+{m_2\mbox{\boldmath $x$}}_2
  +{m_3\mbox{\boldmath $x$}}_3}{m_1+m_2+m_3})
  -(\frac{{m_4\mbox{\boldmath $x$}}_4+{m_5\mbox{\boldmath $x$}}_5}{m_4+m_5}). \nonumber
\end{eqnarray}
$\boldsymbol{x}_i$ is the position of the $i$-th particle. Then we use a set of gaussians to expand the radial part
of the orbital wave function which is shown below,
\begin{eqnarray}
\psi_{lm}(\mathbf{r})=\sum^{n_{max}}_{n=1}c_{nl}\phi^{G}_{nlm}(\mathbf{r})
\end{eqnarray}
\begin{eqnarray}
\phi^{G}_{nlm}(\mathbf{r})=\emph{N}_{nl}r^{l}e^{-\nu_{n}r^{2}}\emph{Y}_{lm}(\hat{\mathbf{r}})
\end{eqnarray}
where $N_{nl}$ is the normalization constant,
\begin{eqnarray}
\emph{N}_{nl}=\left(\frac{2^{l+2}(2\nu_{n})^{l+3/2}}{\sqrt\pi(2l+1)!!}\right)^{\frac{1}{2}},
\end{eqnarray}
and $c_{nl}$ is the variational parameter, which is determined by the dynamics of the system. The Gaussian size
parameters are chosen according to the following geometric progression:
\begin{eqnarray} \label{eq1}
\nu_{n}=\frac{1}{r^{2}_{n}}, r_{n}=r_{min}a^{n-1}, a=\left(\frac{r_{max}}{r_{min}}\right)^{\frac{1}{n_{max}-1}},
\end{eqnarray}
where $n_{max}$ is the number of Gaussian functions, and $n_{max}$ is determined by the convergence of the results.
In the present calculation, $n_{max}=8$ and the results of calculation tends to be stable.

The details of constructing flavor, color and spin wave functions of 5-quark system can be found in Ref.~\cite{32}, and only the last expressions
are shown here.

For flavor wave functions, two possible quantum numbers $I=0$ and $I=1$ in two configurations are considered:
\begin{eqnarray}
 |\chi^{f1}_{0,0} \rangle & = & \frac{1}{\sqrt{2}}(ssu\bar{d}s-ssd\bar{d}s) \nonumber \\
 |\chi^{f2}_{0,0} \rangle & = & -\frac{1}{\sqrt{2}}(sss\bar{u}u+sss\bar{d}d) \nonumber \\
 |\chi^{f3}_{1,1} \rangle & = & ssu\bar{d}s \nonumber \\
 |\chi^{f4}_{1,1} \rangle & = & sss\bar{d}u
\end{eqnarray}

Color wave functions:
\begin{eqnarray}
 |\chi^{c1} \rangle  & = & \frac{1}{\sqrt{18}}(rgb-rbg+gbr-grb+brg-bgr) \nonumber \\
 & & (\bar r r+\bar gg+\bar bb) \nonumber \\
 |\chi^{c2} \rangle  & = &  \frac{1}{\sqrt{192}}\left[ 2(2rrg-rgr-grr)\bar r b \right. \nonumber \\
& + & 2(rgg+grg-2ggr)\bar g b \nonumber \\
& - & 2(2rrb-rbr-brr)\bar r g-2(rbb+brb-2bbr)\bar b g \nonumber \\
& + & 2(2ggb-gbg-bgg)\bar g r+2(gbb+bgb-2bbg)\bar b r \nonumber \\
& + & (rbg-gbr+brg-bgr)(2\bar b b-\bar r r-\bar g g)  \nonumber \\
& + & (2rgb-rbg+2grb-gbr-brg-bgr)(\bar r r-\bar g g)]    \nonumber  \\
 |\chi^{c3} \rangle  & = &  \frac{1}{24}[6(rgr-grr)\bar r b+6(rgg-grg)\bar g b  \\
& - & 6(rbr-brr)\bar r g-6(rbb-brb)\bar b g \nonumber \\
& + & 6(gbg-bgg)\bar g r+6(gbb-bgb)\bar b r \nonumber \\
& + & 3(rbg+gbr-brg-bgr)(\bar r r-\bar g g)  \nonumber \\
& + & (2rgb+rbg-2grb-gbr-brg+bgr)  \nonumber \\
&   & (2\bar b b-\bar r r-\bar g g)]  \nonumber
\end{eqnarray}

Spin wave functions:
\begin{eqnarray}
 |\chi^{\sigma1}_{\frac12,\frac12} \rangle & = & \frac{1}{\sqrt{12}}(2\alpha\alpha\beta\alpha\beta-2\alpha\alpha\beta\beta\alpha
   +\alpha\beta\alpha\beta\alpha \nonumber \\
  & - & \alpha\beta\alpha\alpha\beta+\beta\alpha\alpha\beta\alpha-\beta\alpha\alpha\alpha\beta) \nonumber \\
 |\chi^{\sigma2}_{\frac12,\frac12} \rangle & = & \frac{1}{2}(\alpha\beta\alpha\alpha\beta-\alpha\beta\alpha\beta\alpha+\beta\alpha\alpha\beta\alpha-\beta\alpha\alpha\alpha\beta) \nonumber \\
 |\chi^{\sigma3}_{\frac12,\frac12} \rangle & = & \frac{1}{6}(2\alpha\alpha\beta\alpha\beta+2\alpha\alpha\beta\beta\alpha-\alpha\beta\alpha\alpha\beta-\alpha\beta\alpha\beta\alpha \nonumber \\
& - & \beta\alpha\alpha\beta\alpha-\beta\alpha\alpha\alpha\beta-2\alpha\beta\beta\alpha\alpha-2\beta\alpha\beta\alpha\alpha \nonumber \\
& + & 4\beta\beta\alpha\alpha\alpha) \nonumber \\
 |\chi^{\sigma4}_{\frac12,\frac12} \rangle & = & \frac{1}{\sqrt{12}}(\alpha\beta\alpha\alpha\beta+\alpha\beta\alpha\beta\alpha-\beta\alpha\alpha\alpha\beta-\beta\alpha\alpha\beta\alpha \nonumber \\
& + & 2\beta\alpha\beta\alpha\alpha-2\alpha\beta\beta\alpha\alpha)
\end{eqnarray}
\begin{eqnarray}
 |\chi^{\sigma5}_{\frac12,\frac12} \rangle & = & \frac{1}{\sqrt{18}}(3\alpha\alpha\alpha\beta\beta-\alpha\alpha\beta\alpha\beta
  -\alpha\alpha\beta\beta\alpha \nonumber \\
 & - & \alpha\beta\alpha\alpha\beta-\alpha\beta\alpha\beta\alpha-\beta\alpha\alpha\alpha\beta-\beta\alpha\alpha\beta\alpha  \nonumber \\
& + & \beta\beta\alpha\alpha\alpha+\beta\alpha\beta\alpha\alpha+\alpha\beta\beta\alpha\alpha)  \nonumber \\
 |\chi^{\sigma 6}_{\frac32,\frac32} \rangle & = & \frac{1}{\sqrt{6}}(2\alpha\alpha\beta\alpha\alpha-\alpha\beta\alpha\alpha\alpha
 -\beta\alpha\alpha\alpha\alpha) \nonumber \\
 |\chi^{\sigma 7}_{\frac32,\frac32} \rangle & = & \frac{1}{\sqrt{2}}(\alpha\beta\alpha\alpha\alpha-\beta\alpha\alpha\alpha\alpha) \nonumber \\
 |\chi^{\sigma 8}_{\frac32,\frac32} \rangle & = & \frac{1}{\sqrt{2}}(\alpha\alpha\alpha\alpha\beta-\alpha\alpha\alpha\beta\alpha) \nonumber \\
 |\chi^{\sigma 9}_{\frac32,\frac32} \rangle & = & \frac{1}{\sqrt{30}}(2\alpha\alpha\beta\alpha\alpha+2\alpha\beta\alpha\alpha\alpha+2\beta\alpha\alpha\alpha\alpha \nonumber \\
& - & 3\alpha\alpha\alpha\alpha\beta-3\alpha\alpha\alpha\beta\alpha)  \nonumber\\
 |\chi^{\sigma 10}_{\frac52,\frac52} \rangle & = & \alpha\alpha\alpha\alpha\alpha , \nonumber
\end{eqnarray}
where $\chi^{c1} $ represents the color wave function of a color singlet-singlet structure, $\chi^{c2}$ and $\chi^{c3} $ represent the color octet-octet
wave functions respectively. The subscripts of $\chi^{f}_{I,I_z}$ ($\chi^{\sigma}_{S,S_z}$) are total isospin (spin) and its third projection.

Finally, the total wave function of the five-quark system is written as:
\begin{equation}
\Psi_{JM_J}^{i,j,k}={\cal A} \left[ \left[  \psi_{L}\chi^{\sigma_i}_{S}\right]_{JM_J} \chi^{fi}_j \chi^{ci}_k \right],
\end{equation}
where the $\cal{A}$ is the antisymmetry operator of the system which guarantees the antisymmetry of the total wave functions when identical
particles exchange. Under our numbering scheme, the antisymmetry operator has the following form,
\begin{equation}
 \mathcal{A}  = 1-(15)-(25).  \nonumber \\
\end{equation}
At last, we solve the following Schr\"{o}dinger equation to obtain eigen-energies of the system,
\begin{equation}
H\Psi_{JM_J}=E\Psi_{JM_J},
\end{equation}
with the help of the Rayleigh-Ritz variational principle. The matrix elements of Hamiltonian can be easily obtained if all the orbital angular momenta
are zero, which is reasonable for only considering the low-lying states of pentaquark systems. It is worthwhile to mention that if the orbital angular
momenta of the systems are not zero, it is necessary to use the infinitesimally shifted Gaussian method to calculate the matrix elements~\cite{31}.

\section{Results and discussions}
In this section, we present the results of possible $S$-wave of pentaquark systems $ssq\bar qs,q=u,d$ with all possible quantum numbers
$IJ^P=0(\frac{1}{2})^-$, $0(\frac{3}{2})^-$, $0(\frac{5}{2})^-$, $1(\frac{1}{2})^-$, $1(\frac{3}{2})^-$ and $1(\frac{5}{2})^-$.
All the orbital angular momentum of the systems are treated as zero and the corresponding parity is negative.

\begin{table*}[htb]
\caption{The results for system with $IJ^P=0\frac{1}{2}^-$. cc1: mixing of color singlet-singlet channels,
cc2: mixing of all channels. (unit: MeV)\label{e012}}
\begin{tabular}{cccccccccc}
\hline \hline
~~~~~$\psi^{f_i}$~~~~~~&~~~~~$\psi^{\sigma_j}$~~~~~&~~~~~$\psi^{c_k}$~~~~~&~~~~Channel~~~~~&~~~E~~~~&~~~$E^{Theo}_{th}$~~~~&
  ~~~$E_B$~~~&~~~$E^{Exp}_{th}$~~~&~~~$E^{\prime}$~~~\\
\hline
 $i=1$ & $j=1$ & $k=1$ & $\Xi \bar{K}$ & 1816 & 1816  &0 & 1813 & 1813 \\
 $i=1$ & $j=1,2$ & $k=1,2,3$ & & 1816 & &  & &  \\
 $i=1$ & $j=3$ & $k=1$ & $\Xi \bar{K}^*$ & 2178 & 2232  &-54 & 2207 & 2153 \\
 $i=1$ & $j=3,4$ & $k=1,2,3$ & & 2166 & & -66 & & 2141 \\
 $i=1$ & $j=5$ & $k=1$ & $\Xi^*\bar{K}^*$ &2438 & 2451  & -13 & 2422 & 2409 \\
 $i=1$ & $j=5$ & $k=1,3$ & & 2430 & & -21 & & 2401 \\
  $i=2$ & $j=5$ & $k=1$ & $\Omega\omega$ &2261 & 2262  & 0 & 0 & 2454 \\
 $i=2$ & $j=5$ & $k=1,3$ & & 2261 & &  & &  \\
cc1 & & & & & 1816 &  0&   \\
cc2 & & & & & 1816  \\
\hline \hline
\end{tabular}
\end{table*}

\begin{table*}[htb]
\caption{The results for system with $IJ^P=1\frac{1}{2}^-$. cc1: mixing of color singlet-singlet channels,
cc2: mixing of all channels. (unit: MeV)\label{e112}}
\begin{tabular}{cccccccccc}
\hline \hline
~~~~~$\psi^{f_i}$~~~~~~&~~~~~$\psi^{\sigma_j}$~~~~~&~~~~~$\psi^{c_k}$~~~~~&~~~~Channel~~~~~&~~~E~~~~&~~~$E^{Theo}_{th}$
 ~~~~&~~~$E_B$~~~&~~~$E^{Exp}_{th}$~~~&~~~$E^{\prime}$~~~\\
\hline
 $i=3$ & $j=1$ & $k=1$ & $\Xi \bar{K}$ & 1816 & 1816  &0 & 1813 & 1813 \\
 $i=3$ & $j=1,2$ & $k=1,2,3$ & & 1816 & &  & &  \\
 $i=3$ & $j=3$ & $k=1$ & $\Xi \bar{K}^*$ & 2220 & 2232  &-12 & 2207 & 2195 \\
 $i=3$ & $j=3,4$ & $k=1,2,3$ & & 2208 & & -24 & & 2283 \\
 $i=3$ & $j=5$ & $k=1$ & $\Xi^*\bar{K}^*$ &2431 & 2451  & -20 & 2422 & 2402 \\
 $i=3$ & $j=5$ & $k=1,3$ & & 2420 & & -31 & & 2391 \\
  $i=4$ & $j=5$ & $k=1$ & $\Omega\rho$ &2440 & 2440  & 0 & 0 & 2447 \\
 $i=4$ & $j=5$ & $k=1,3$ & & 2440 & &  & &  \\
cc1 & & & & & 1816 &  0&   \\
cc2 & & & & & 1816  \\
\hline \hline
\end{tabular}
\end{table*}

\begin{table*}[htb]
\caption{The results for system with $IJ^P=0\frac{3}{2}^-$. cc1: mixing of color singlet-singlet channels,
cc2: mixing of all channels. (unit: MeV)\label{e032}}
\begin{tabular}{cccccccccc}
\hline \hline
~~~~~$\psi^{f_i}$~~~~~~&~~~~~$\psi^{\sigma_j}$~~~~~&~~~~~$\psi^{c_k}$~~~~~&~~~~Channel~~~~~&~~~E~~~~&~~~$E^{Theo}_{th}$~~~~&
  ~~~$E_B$~~~&~~~$E^{Exp}_{th}$~~~&~~~$E^{\prime}$~~~\\
\hline
 $i=1$ & $j=6$ & $k=1$ & $\Xi \bar{K}^*$ & 2232 & 2232  &0 & 2207 & 2207 \\
 $i=1$ & $j=6,7$ & $k=1,2,3$ & & 2232 & &  & &  \\
 $i=1$ & $j=8$ & $k=1$ & $\Xi^*\bar{K}$ & 2035 & 2035  &0 & 2028 & 2028 \\
 $i=1$ & $j=8$ & $k=1,3$ & & 2035 & &  & &  \\
 $i=1$ & $j=9$ & $k=1$ & $\Xi^*\bar{K}^*$ &2413 & 2451  & -38 & 2422 & 2384 \\
 $i=1$ & $j=9$ & $k=1,3$ & & 2378 & & -73 & & 2349 \\
  $i=2$ & $j=8$ & $k=1$ & $\Omega\eta$ &2232 & 2232  & 0 & 0 & 2220 \\
 $i=2$ & $j=8$ & $k=1,3$ & & 2232 & &  & &  \\
 $i=2$ & $j=9$ & $k=1$ & $\Omega\omega$ &2262 & 2262  & 0 & 0 & 2454 \\
 $i=2$ & $j=9$ & $k=1,3$ & & 2262 & &  & &  \\

cc1 & & & & & 2029 &  -6&  \\
cc2 & & & & & 2025 & -10& \\
\hline \hline
\end{tabular}
\end{table*}

\begin{table*}[htb]
\caption{The results for system with $IJ^P=1\frac{3}{2}^-$. cc1: mixing of color singlet-singlet channels,
 cc2: mixing of all channels. (unit: MeV)\label{e132}}
\begin{tabular}{cccccccccc}
\hline \hline
~~~~~$\psi^{f_i}$~~~~~~&~~~~~$\psi^{\sigma_j}$~~~~~&~~~~~$\psi^{c_k}$~~~~~&~~~~Channel~~~~~&~~~E~~~~&~~~$E^{Theo}_{th}$~~~~&
 ~~~$E_B$~~~&~~~$E^{Exp}_{th}$~~~&~~~$E^{\prime}$~~~\\
\hline
 $i=3$ & $j=6$ & $k=1$ & $\Xi \bar{K}^*$ & 2232 & 2232  &0 & 2207 & 2207 \\
 $i=3$ & $j=6,7$ & $k=1,2,3$ & & 2232 & &  & &  \\
 $i=3$ & $j=8$ & $k=1$ & $\Xi^*\bar{K}$ & 2035 & 2035  &0 & 2028 & 2028 \\
 $i=3$ & $j=8$ & $k=1,3$ & & 2035 & &  & &  \\
 $i=3$ & $j=9$ & $k=1$ & $\Xi^*\bar{K}^*$ &2444 & 2451  & -7 & 2422 & 2415 \\
 $i=3$ & $j=9$ & $k=1,3$ & & 2436 & & -15 & & 2407 \\
  $i=4$ & $j=8$ & $k=1$ & $\Omega\pi$ &1811 & 1811  & 0 & 1812 & 1812 \\
 $i=4$ & $j=8$ & $k=1,3$ & & 1811 & &  & &  \\
 $i=4$ & $j=9$ & $k=1$ & $\Omega\rho$ &2440 & 2440  & 0 & 0 & 2447 \\
 $i=4$ & $j=9$ & $k=1,3$ & & 2440 & &  & &  \\

cc1 & & & & & 1810 &  -1&  \\
cc2 & & & & & 1809 & -2& \\
\hline \hline
\end{tabular}
\end{table*}

\begin{table*}[htb]
\caption{The results for system with $IJ^P=0\frac{5}{2}^-$. cc1: mixing of color singlet-singlet channels,
cc2: mixing of all channels. (unit: MeV)\label{e052}}
\begin{tabular}{cccccccccc}
\hline \hline
~~~~~$\psi^{f_i}$~~~~~~&~~~~~$\psi^{\sigma_j}$~~~~~&~~~~~$\psi^{c_k}$~~~~~&~~~~Channel~~~~~&~~~E~~~~&~~~$E^{Theo}_{th}$~~~~&
 ~~~$E_B$~~~&~~~$E^{Exp}_{th}$~~~&~~~$E^{\prime}$~~~\\
\hline
 $i=1$ & $j=10$ & $k=1$ & $\Xi^*\bar{K}^*$ & 2451 & 2451  &0 & 2422 & 2422 \\
 $i=1$ & $j=10$ & $k=1,3$ & & 2451 & &  & &  \\
 $i=2$ & $j=10$ & $k=1$ & $\Omega\omega$ & 2262 & 2262  &0 & 2454 & 2454 \\
 $i=2$ & $j=10$ & $k=1,3$ & & 2262 & &  & &  \\
cc1 & & & & & 2262 &  0&  \\
cc2 & & & & & 2262 & & \\
\hline \hline
\end{tabular}
\end{table*}

\begin{table*}[htb]
\caption{The results for system with $IJ^P=1\frac{5}{2}^-$. cc1: mixing of color singlet-singlet channels,
cc2: mixing of all channels. (unit: MeV)\label{e152}}
\begin{tabular}{cccccccccc}
\hline \hline
~~~~~$\psi^{f_i}$~~~~~~&~~~~~$\psi^{\sigma_j}$~~~~~&~~~~~$\psi^{c_k}$~~~~~&~~~~Channel~~~~~&~~~E~~~~&~~~$E^{Theo}_{th}$~~~~&
~~~$E_B$~~~&~~~$E^{Exp}_{th}$~~~&~~~$E^{\prime}$~~~\\
\hline
 $i=3$ & $j=10$ & $k=1$ & $\Xi^*\bar{K}^*$ & 2451 & 2451  &0 & 2422 & 2422 \\
 $i=3$ & $j=10$ & $k=1,3$ & & 2447 & &  & &  \\
 $i=4$ & $j=10$ & $k=1$ & $\Omega\rho$ & 2440 & 2440  &0 & 2447 & 2447 \\
 $i=4$ & $j=10$ & $k=1,3$ & & 2440 & &  & &  \\
cc1 & & & & & 2440 &  0&  \\
cc2 & & & & & 2440 & & \\
\hline \hline
\end{tabular}
\end{table*}

In Tables \ref{e012}-\ref{e152}, the significant calculation results are shown. In each table, columns 1 to 3 represent the indices of flavor, spin and color
wave functions in each channel. Column 4 is the corresponding physical channel of pentaquark systems. In column 5, the eigen-energy of each channel
is listed and the theoretical threshold (the sum of the theoretical masses of corresponding baryon and meson) is given in column 6. Column 8 gives
the binding energies, which are the difference between the eigen-energy and the theoretical threshold. Finally, the experimental thresholds
(the sum of the experimental masses of the corresponding baryon and meson) along with corrected energies (the sum of experimental threshold and
the binding energy, $E^{\prime}=E_B+E_{th}^{exp}$) are given in last two columns. With this correction, the calculation error caused by the model
parameters in pentaquark calculation can be reduced partly.  The results are analyzed in the following.

(a) $IJ^P=0\frac{1}{2}^-$ (Table \ref{e012}): Single-color structure calculation shows that there exist two bound states in 
$\Xi \bar{K}^*$ and $\Xi^* \bar{K}^*$ channels while $\Xi \bar{K}$ and $\Omega\eta$ are unbound. After coupling to respective hidden-color channels, 
the binding energy of two bound states increase by a few MeVs. When coupled-channel calculation between both single-color structure and all type of
structure is performed, the lowest energy is above the threshold of $\Xi \bar{K}$. So no bound state can be formed in $IJ^P=0\frac{1}{2}^-$
system. However, resonances are possible because the strong attractions exist both in $\Xi \bar{K}^*$ and $\Xi^* \bar{K}^*$ channels.
Thus, real-scaling method is required to identify resonances.

(b) $IJ^P=1\frac{1}{2}^-$ (Table \ref{e112}): The results are similar to that of $IJ^P=0\frac{1}{2}^-$ system, there are bound states in 
$\Xi \bar{K}^*$ and $\Xi^* \bar{K}^*$ channels and other two channels are unbound. The lowest eigen-energy of channel coupling calculation is above the
threshold of $\Xi \bar{K}$, so no bound state found in $IJ^P=1\frac{1}{2}^-$ system. $\Xi \bar{K}^*$ and $\Xi^* \bar{K}^*$ are possible resonances.

(c) $IJ^P=0\frac{3}{2}^-$ (Table \ref{e032}): The single channel calculations show that there exists bound state in $\Xi^* \bar{K}^*$ channel, 
Moreover, the attraction increases after coupling to its hidden-color channel. Although the single channel calculation shows that the state $\Xi^* \bar{K}$ 
is unbound, the couple-channels calculations push the states below the threshold. The coupling of all color singlet-singlet channels find the lowest
eigen-energy of the system is 2029 MeV, 6 MeV below the threshold, The full channel coupling lower the lowest eigen-energy to 2025 MeV.
So a bound state (main component is $\Xi^* \bar{K}$) can be formed after the channel coupling in $IJ^P=0\frac{3}{2}^-$ system, which is regarded as
a good candidate for $\Omega(2012)$. Also, it is possible for $\Xi^* \bar{K}^*$ to form a resonance.

(d) $IJ^P=1\frac{3}{2}^-$ (Table \ref{e132}): The results are again similar to that of $IJ^P=0\frac{3}{2}^-$ system. There is only one bound state,
$\Xi^* \bar{K}^*$ in the single channel calculation, and it is possible to become a resonance. In addition, the channel coupling leads to a bound state
in the system. the lowest eigen-energy is 1810 MeV (color singlet-singlet channel coupling), or 1809 MeV (full channel coupling), 1 MeV or 2 MeV
lower than the threshold of the lowest channel $\Omega\pi$. So there is a bound state for the $IJ^P=1\frac{3}{2}^-$ system.

(e) For $J^P=\frac{5}{2}^-,I=0,1$ (Table \ref{e052} and (Table \ref{e152}): The results of these two systems are similar. There is no bound state and 
the conclusions are the same after full-channel coupling.

From the above results, one can that the iso-scalar and iso-vector states have similar behavior, generally the iso-vector states have a little higher
energy than that of the corresponding iso-scalar states. This fact infers that the state are molecular one, the large separation between baryon and
meson minimize the contribution from the pion exchange potential. In the following, we shall calculation the separations between quarks to verify the 
molecule picture of the state.

We make a comparison between our calculation results and that of Ref.~\cite{30}. The binding energies of states obtained in our work are generally bigger, 
basically on the order of tens since we consider the effects of scalar nonet exchange, whose contributions are strongly attractive to bind two hadrons 
together in our work, in Ref.~\cite{30}, only $\sigma$-exchange between non-strange quarks is considered. For example, the binding energy of 
$[\Xi \bar{K}^*]_{0\frac{1}{2}^-}$ is 54 MeV, whereas it is 8 MeV in Ref.~\cite{30}.
And it is reasonable that $[\Xi^* \bar{K}^*]_{0\frac{1}{2}^-}$, $[\Xi \bar{K}^*]_{1\frac{1}{2}^-}$ and $[\Xi^*\bar{K}^*]_{1\frac{1}{2}^-}$ with more than 
a dozen of binding energies in our work and they become unbound in Ref.~\cite{30}. For $IJ^P=0\frac{3}{2}^-$ system, we solve the mass inversion problem 
of $\Xi^* \bar{K}$ and $\Omega\eta$, which leads to a more objective description on the effect of channel-coupling. Then, come to $IJ^P=1\frac{3}{2}^-$ system, 
there is no mass inversion problem in both two calculations, we found that our results are consistent with that of Ref.~\cite{30}. Finally, 
for $IJ^P=0\frac{5}{2}^-$ and $1\frac{5}{2}^-$ systems, the results are the same in both two calculations, there is no any bound state.

To check whether the resonances in $IJ^P=0\frac{1}{2}^-$, $0\frac{3}{2}^-$, $1\frac{3}{2}^-$ and $1\frac{3}{2}^-$ systems can survive after coupling to 
the open channels, a stability method to identify genuine resonance states, real-scaling method (stabilization method)~\cite{36,37,38}, is employed. 
In this method, the Gaussian size parameter $r_{n}$, appearing in Eq.~(\ref{eq1}), for the basis functions between the color-singlet baryon and meson
clusters is scaled by a factor $\alpha$: $r_{n} \longrightarrow \alpha r_{n}$. As a result, a genuine resonance will act as an avoid-crossing structure
(see Fig.~1) with the increasing of $\alpha$, while other continuum states will fall off towards its threshold. If the avoid-crossing structure is repeated periodically as $\alpha$ increase, then the avoid-crossing structure is a genuine resonance.
\begin{figure}[h]
  \centering
  \includegraphics[width=9cm,height=7cm]{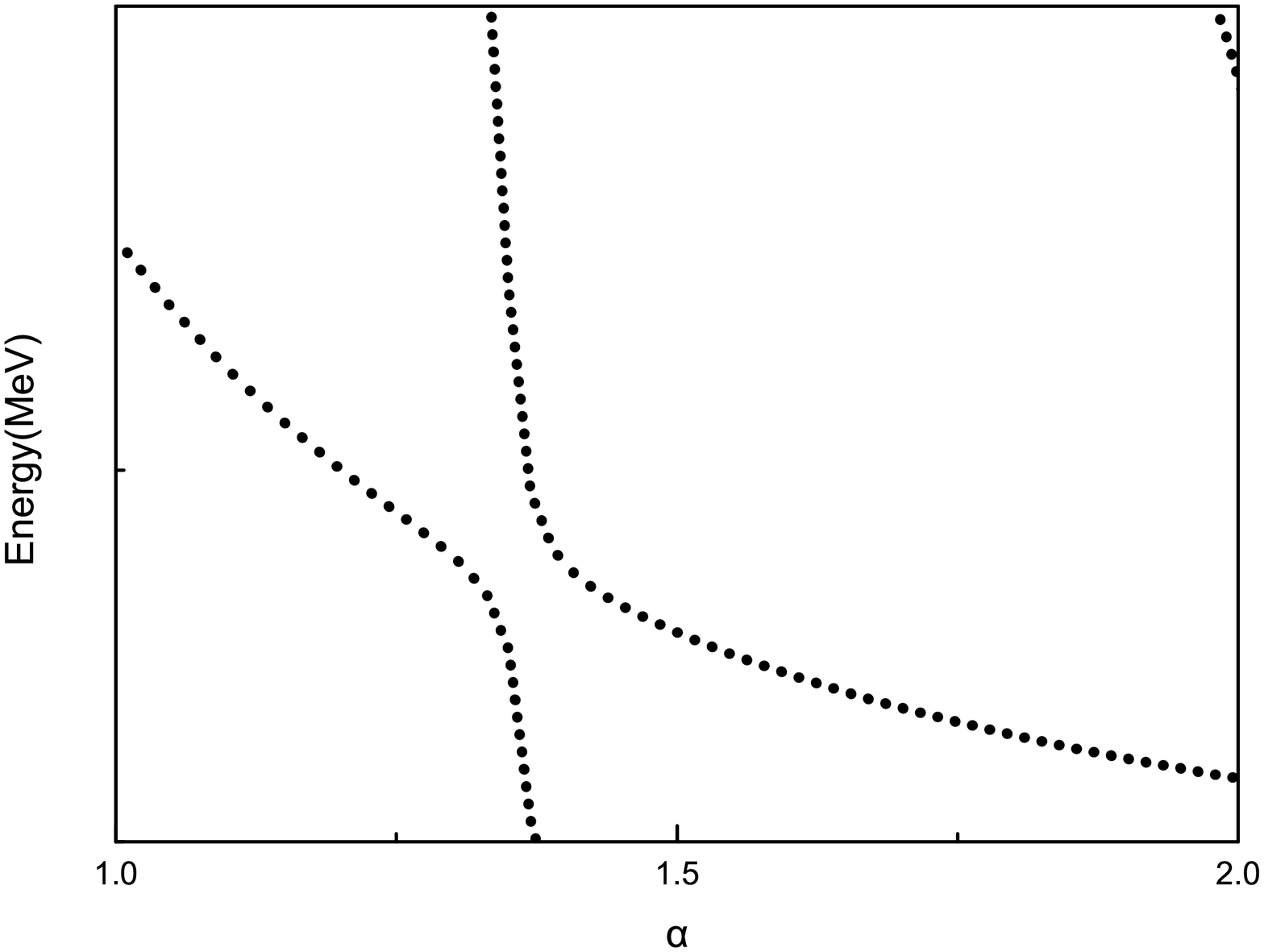}
  \caption{The shape of the resonance in real-scaling method}
  \label{1}
\end{figure}

Our results are shown in Figs.~2, 3, 4 and 5. In these figures, the thresholds of all physical channels appear as horizontal lines and are marked with lines 
(red lines), along with their contents tagged. And for genuine resonances, which appear as avoid-crossing structure and are marked with blue lines. 
Bound states are also marked with blue line below the lowest threshold of the systems. The continuum states fall off towards their respective threshold 
(red horizontal lines). For $IJ^P=0\frac{5}{2}^-, 1\frac{5}{2}^-$ systems, since there is no any bound state in single-channel calculations, which means no 
mechanism of forming resonances, we do not present the results of these two systems.

\begin{figure}[h]
  \centering
  \includegraphics[width=9cm,height=7cm]{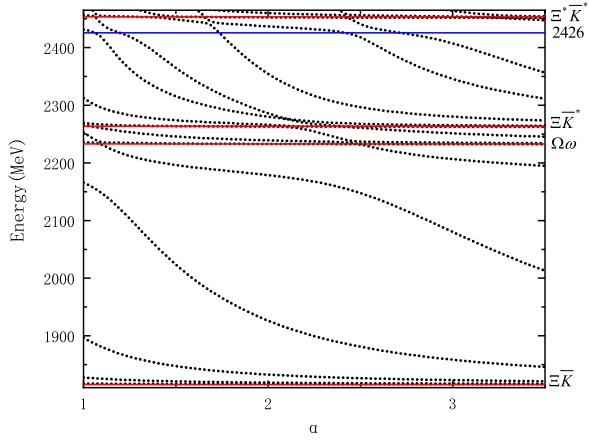}
  \caption{Energy spectrum of $0\frac{1}{2}^-$ system.}
  \label{1}
\end{figure}
\begin{figure}[h]
  \centering
  \includegraphics[width=9cm,height=7cm]{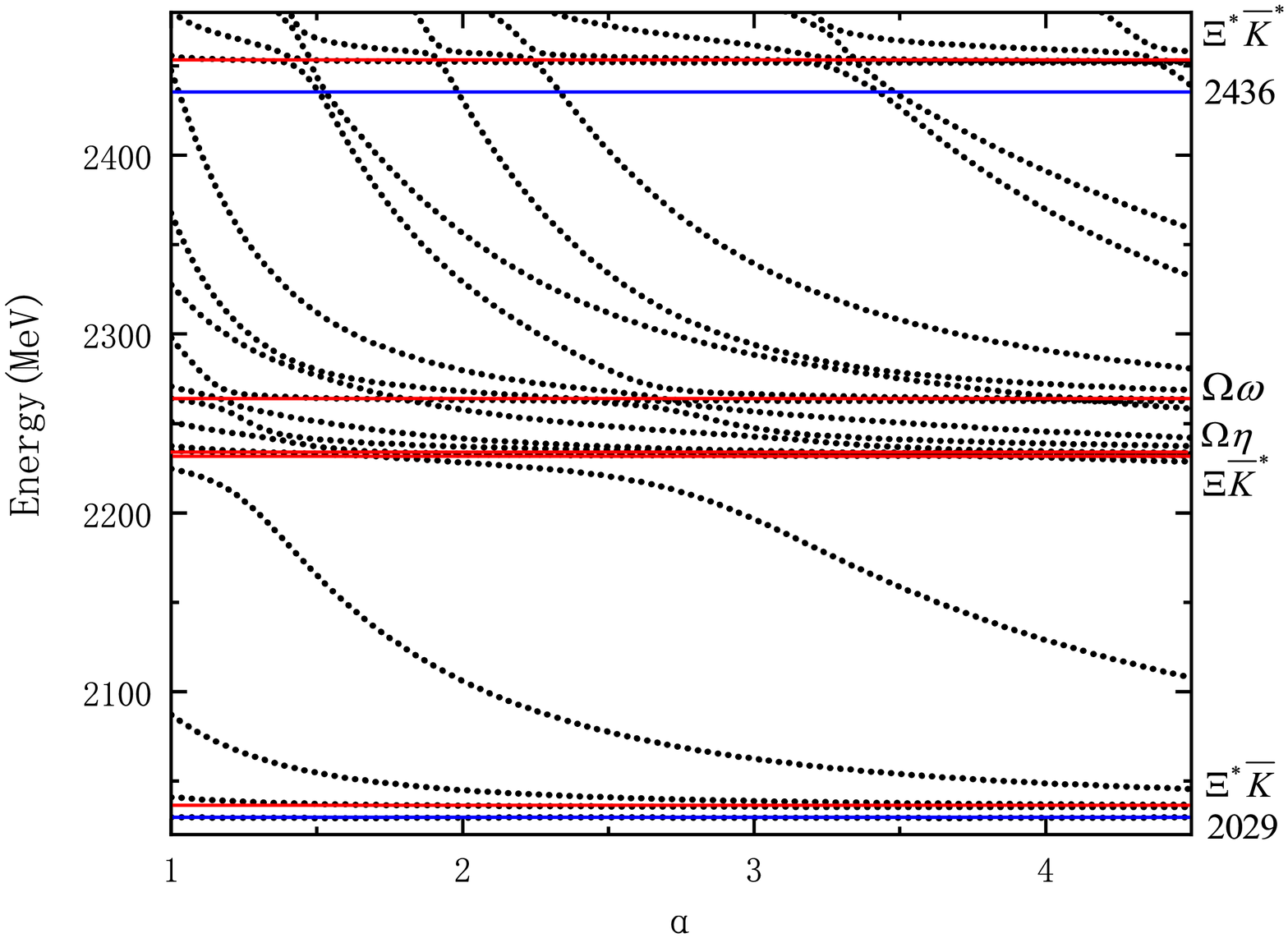}
  \caption{Energy spectrum of $0\frac{3}{2}^-$ system.}
  \label{1}
\end{figure}
\begin{figure}[h]
  \centering
  \includegraphics[width=9cm,height=7cm]{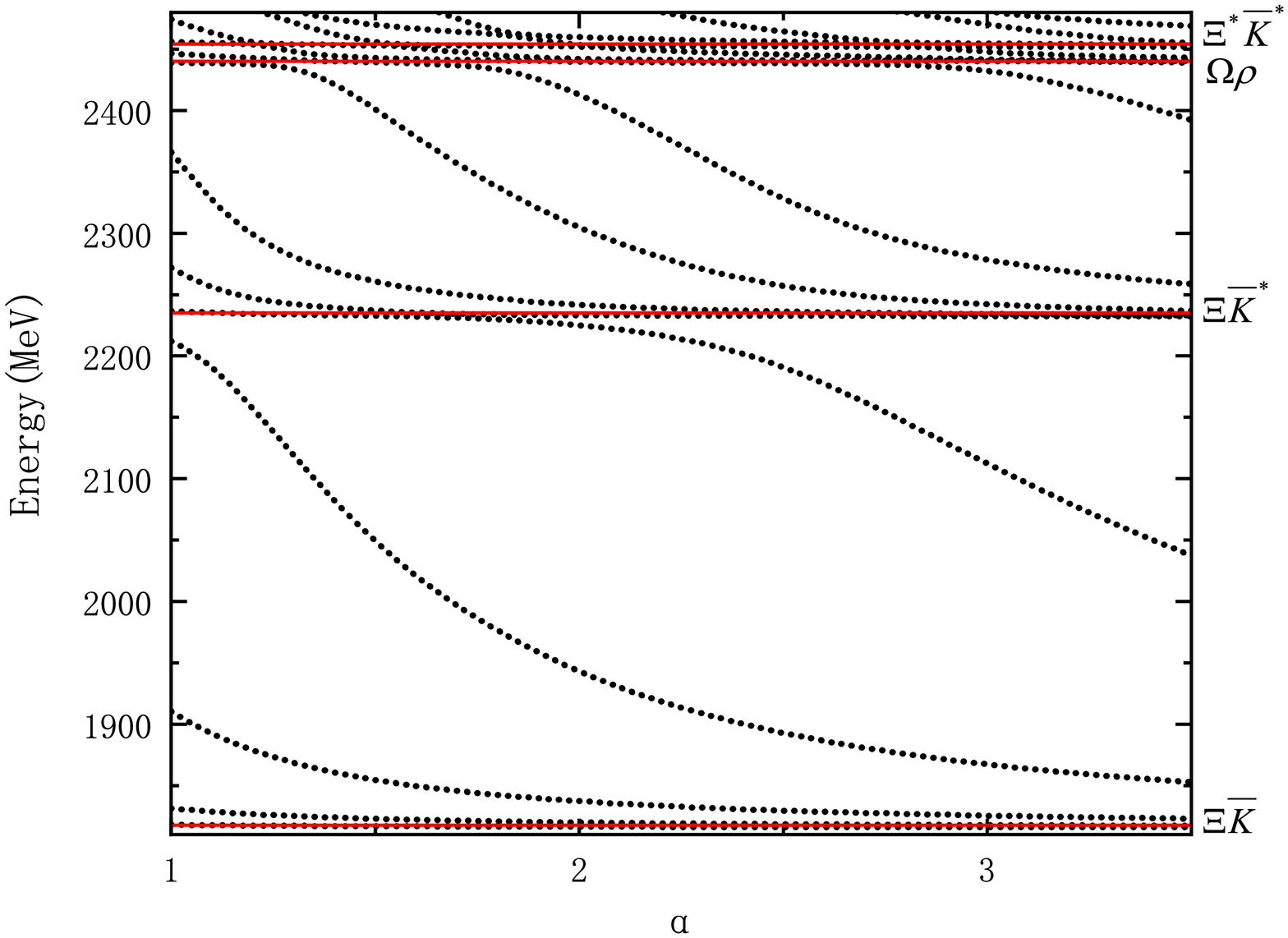}
  \caption{Energy spectrum of $1\frac{1}{2}^-$ system.}
  \label{1}
\end{figure}
\begin{figure}[h]
  \centering
  \includegraphics[width=9cm,height=7cm]{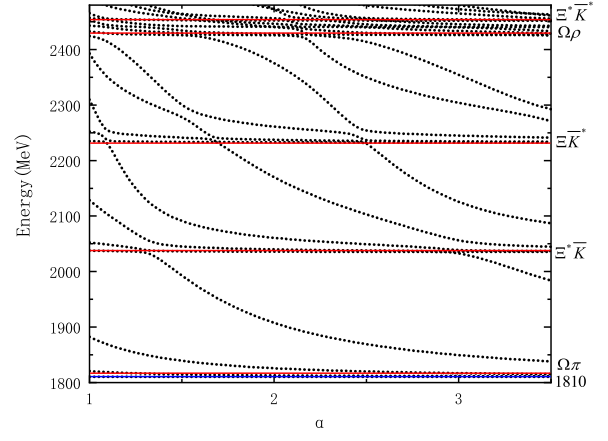}
  \caption{Energy spectrum of $1\frac{3}{2}^-$ system.}
  \label{1}
\end{figure}

For $IJ^P=0\frac{1}{2}^-$ system, we get one resonance whose energy is 2426 MeV (main component is $\Xi^* \bar{K}^*$). The other possible resonance, 
$\Xi \bar{K}^*$, decays strongly to the open channels and cannot form an avoid-crossing structure, so we do not think it is a observable resonance. 
In $IJ^P=0\frac{3}{2}^-$ system, there is only one resonance and its energy is 2436 MeV (main component is $\Xi^* \bar{K}^*$), and there is also a bound state
which lies below the lowest threshold (a blue line marked by 2029). 
Although there are two channels where exist weakly bound states in the single channel calculation for the $IJ^P=1\frac{1}{2}^-$ system, we found no 
avoid-crossing structure, which means there is no any resonance. Finally, in $IJ^P=1\frac{3}{2}^-$ system, there is no any resonance. 
However, the lowest energy of this system is pushed below the lowest threshold of the system $\Omega\pi$ by the channel coupling, 
thus there is a bound states which is shown as a blue line marked by 1810. It is worth mentioning that in case we use real-scaling method to identify 
resonances, we will also calculate the composition of the possible resonances to find the mechanism of the formation of the resonances. 
The main component of a genuine resonance should be bound state channels in the single channel calculation or coupled channels with energy below the 
the lowest threshold of the coupling channels in the channel coupling calculation. For example, in Fig.~5, there is an avoid-crossing structure 
around 2250 MeV. However, its main component is $\Omega\omega$, a open channel, which means the avoid-crossing structure is formed by two or more open 
channels which have different falling slopes.

\begin{table*}[th]
\caption{The decay width and RSM distances of $\Omega$ states in all systems.\label{size}}
\begin{tabular}{ccccccccccccccccccccccccccccccccccccccccccccccccc} \hline \hline
Model&~~$IJ^P$~~  &State&~~   &Main comp.&~~  &$E'$&~~  &Width&~~   &$r_{12}$&~ &$r_{13}$&~  &$r_{14}$&~  &$r_{15}$&    &$r_{34}$&  &$r_{35}$&  &$r_{45}$&\\ \hline
Resonance &$0\frac{1}{2}^-$  &$\Omega(2426)$&  &$\Xi^* \bar{K}^*$ &   &2396&  &73.5& &0.7& &0.9&  &1.5& &1.6&  &1.6&  &1.6&  &0.9&\\
          &$0\frac{3}{2}^-$  &$\Omega(2436)$&  &$\Xi^* \bar{K}^*$ &   &2417&  &68.4& &0.5& &0.9&  &2.1& &2.2&  &2.2&  &2.1&  &0.8&\\ \hline \hline
Bound state & $0\frac{3}{2}^-$ &$\Omega(2029)$&&$\Xi^* \bar{K}$ &     &2018&  &-& &0.7& &0.8&  &2.3& &2.3&  &2.3&  &2.3&  &0.5&\\
            &$1\frac{3}{2}^-$  &$\Omega(1809)$&&$\Omega\pi$ &         &1810&  &-& &0.7& &0.7&  &3.5& &3.5&  &3.5&  &3.5&  &0.5&\\  \hline \hline
\end{tabular}
\end{table*}

For resonances, the partial widths of two hadrons strong decay can be extracted from these figures. The decay width of resonance to possible open channels 
of two hadrons is obtained by the following formula
\begin{eqnarray}
&&\Gamma=4V(\alpha)\frac{\sqrt{k_r  k_c}}{|k_r-k_c|},
\end{eqnarray}
where $V(\alpha)$ is the minimum energy difference, while $k_c$ and $k_r$ stand for the slopes of scattering state and resonance state respectively. 
More details can be found in Ref.~\cite{36}. And the total widths also include the decay width of each hadron in the pentaquark state. 
For bound states, they have no widths due to their stability against the strong decay. After a series of calculations and identification, we also calculate 
the root-mean-square (RMS) distances between any two quarks in all bound states and observable resonances to determine their inner structure respectively 
and the results are shown in Table \ref{size}. The main component of each resonance and corrected energy are also given.

From Table IX, we can see that, for two resonances $\Omega(2426)$ and $\Omega(2436)$, the decay widths are 73.5 MeV and 68.4 MeV respectively. The distances 
among quarks 1, 2 and 3 are 0.5-0.9 fm and the distances between quark 4 and anti-quark are 0.8-0.9 fm, while the distances between quarks 1, 2, 3 and 4, 5 
are over 1.5 fm. Thus it is natural to describe these two resonances as molecular states. In addition, the distance of $\Omega(2426)$ between two clusters 
is obviously lower than that of $\Omega(2436)$ due to its higher binding energy of main component channel $\Xi^* \bar{K}^*$. 
For $\Omega(2029)$, the distances between quarks shows its structure is molecular state. Also, the binding energy of the state in single channel is 6 MeV 
and after coupling to hidden-color channels, the binding energy increase by a few MeVs, which is the typical range of binding energy of hadronic molecules. 
The corrected energy of molecular state $\Omega(2029)$ is 2018 MeV, closed to 2012 MeV, so it is a good candidate for $\Omega(2012)$ reported 
by Belle collaboration. Nevertheless, in present work, we only consider all possible low-lying states of pentaquark systems with all orbital angular momentum 
set to 0. Release this constraint, the state $\Omega(2029)$ can decay to $D$-waves $\Xi\bar{K}$ via the tensor interaction. So considering the effect of all 
possible $D$-wave channels is our further study. For $\Omega(1809)$, its result is similar to $\Omega(2029)$ and the structure of the state is molecular one. However, the very small binding energy and the distance between two clusters of $\Omega(1809)$ is over 3 fm, which means it is a loose bound state.

\section{Summary}
In this work, all low-lying states of $sssq \bar{q},q=u,d$ pentaquark systems are systematically investigated by means of the real scaling method in the 
framework of chiral quark model, with the help of a high-precision numerical approach, Gaussian expansion method.

The results show that $\Omega(2012)$ reported by Belle Collaboration can be described as $\Xi^* \bar{K}$ molecular state with $IJ^P=0\frac{3}{2}^-$. 
In addition, a very loose bound state $\Omega(1810)$ with $IJ^P=1\frac{3}{2}^-$ and two observable resonances, $\Omega(2396)$, $\Omega(2417)$ with $IJ^P=0\frac{1}{2}^-, 0\frac{3}{2}^-$ are found. The root-mean-square (RMS) distances between any two quarks of these states are also calculated 
to determine their structures, along with their decay widths. The results shows that four stable states are all molecular states. We hope the states 
proposed above can be searched in future experiments. Nevertheless, we cannot ignore the mixing of $S$- and $D$-wave of all states when the spin-orbit 
and tensor interactions are considered. So further study is expected.

\section*{Acknowledgments}
The work is supported partly by the National Natural Science Foundation of China under Grant Nos. 11775118, and 11535005 and 
Postgraduate Research and Practice Innovation Program of Jiangsu Province under Grant No. KYCX22\_1542.


\begin{thebibliography}{99}
\bibitem{1} J. Yelton et al. (Belle Collaboration), Phys. Rev. Lett. 121, 052003 (2018).
\bibitem{2} M. Gell-Mann, Phys. Rev. 125, 1067 (1962).
\bibitem{3} Y. Ne'eman, Nucl. Phys. 26, 222 (1961).
\bibitem{4} N. Isgur and G. Karl, Phys. Rev. D 18, 4187 (1978).
\bibitem{5} S. Capstick and N. Isgur, Phys. Rev. D 34, 2809 (1986); AIP Conf. Proc. 132, 267 (1985).
\bibitem{6} Y. Oh, Phys. Rev. D 75, 074002 (2007).
\bibitem{7} G. P. Engel et al. (BGR Collaboration), Phys. Rev. D 87, 074504 (2013).
\bibitem{8} J.L. Goity, C. Schat, N.N. Scoccola, Phys. Lett. B 564, 83 (2003).
\bibitem{9} L. Y. Xiao and X. H. Zhong, Phys. Rev. D 98, 034004 (2018).
\bibitem{10} T. M. Aliev, K. Azizi, Y. Sarac, and H. Sundu, Phys. Rev. D 98, 014031 (2018).
\bibitem{11} Z.Y. Wang, L.C. Gui, Q.F. L\"{u}, L.Y. Xiao, and X.H. Zhong, Phys. Rev. D 98, 114023 (2018).
\bibitem{12} E. E. Kolomeitsev and M. F. M. Lutz, Phys. Lett. B 585, 243 (2004).
\bibitem{13} S. Sarkar, E. Oset, and M. J. Vicente Vacas, Nucl. Phys. A 750, 294 (2005); 780, 90(E) (2006).
\bibitem{14} S.G. Yuan, C.S. An, K.W. Wei, B.S. Zou, and H.S. Xu, Phys. Rev. C 87 (2013) 025205.
\bibitem{15} S.Q. Xu, J.J. Xie, X.R. Chen, and D.J. Jia, Commun. Theor. Phys. 65, 53 (2016).
\bibitem{16} M.V. Polyakov, H.D. Son, B.D. Sun, A. Tandogan, Phys. Lett. B 792, 315 (2019).
\bibitem{17} F.K. Guo, C. Hanhart, U.G. Meissner, Q. Wang, Q. Zhao, B.S. Zou, Rev. Mod. Phys. 90, 015004 (2018).
\bibitem{18} M. P. Valderrama, Phys. Rev. D 98, 054009 (2018).
\bibitem{19} R. Pavao and E. Oset, Eur. Phys. J. C 78, 857 (2018).
\bibitem{20} Y. H. Lin and B. S. Zou, Phys. Rev. D 98, 056013 (2018).
\bibitem{21} Y. Huang, M. Z. Liu, J. X. Lu, J. J. Xie, and L. S. Geng, Phys. Rev. D 98, 076012 (2018).
\bibitem{22} S. Jia et al. [Belle Collaboration], Phys. Rev. D 100, 032006 (2019).
\bibitem{23} N. Ikeno, G. Toledo, and E. Oset, Phys. Rev. D 101, 094016 (2020).
\bibitem{24} J. X. Lu, C. H. Zeng, E. Wang, J. J. Xie, and L. S. Geng, Eur. Phys. J. C 80, 361 (2020).
\bibitem{25} Y. H. Lin, F. Wang, and B. S. Zou, Phys. Rev. D 102, 074025 (2020).
\bibitem{26} T. Gutsche and V. E. Lyubovitskij, J. Phys. G 48, 025001 (2020)
\bibitem{27} M. S. Liu, K. L. Wang, Q. F. Lu, and X. H. Zhong, Phys. Rev. D 101, 016002 (2020).
\bibitem{28} C. H. Zeng, J. X. Lu, E. Wang, J. J. Xie, and L. S. Geng, Phys. Rev. D 102, 076009 (2020).
\bibitem{29} (Belle Collaboration),  [arXiv:2207.03090 [hep-ex]].
\bibitem{30} X. J. Liu, H. X. Huang, J. L. Ping, and D. Y. Chen, Phys. Rev. C 103, 025202 (2021).
\bibitem{31} E. Hiyama, Y. Kino, and M. Kamimura, Prog. Part. Nucl. Phys. 51, 223 (2003).
\bibitem{32} X. H. Hu, Y. Tan, and J. L. Ping, Eur. Phys. J. C 81, 370 (2021).
\bibitem{33} X. H. Hu, and J. L. Ping, Eur. Phys. J. C 82, 118 (2022).
\bibitem{34} G. Yang, J. L. Ping, and J. Segovia, Phys. Rev. D 102, 054023 (2020).
\bibitem{35} Y. Tan, J. L. Ping, Chin. Phys. C 45, 093104 (2021).
\bibitem{36} J. Simons, J. Chem. Phys. 75, 2465 (1981).
\bibitem{37} E. Hiyama, A. Hosaka, M. Oka, J.M. Richard, Phys. Rev. C 98, 045208 (2018).
\bibitem{38} Q. Meng, E. Hiyama, K.U. Can, P. Gubler, M. Oka, A. Hosaka, H. Zong, Phys. Lett. B 798, 135028 (2019).
\bibitem{39} A. Valcarce, H. Garcilazo, F. Fernandez, and P. Gonzalez, Rep. Prog. Phys. 68, 965 (2005).
\bibitem{40} H. Garcilazo, T. Fernandez-Carames and A. Valcarce, Phys. Rev. C 75, 034002 (2007)
\end{thebibliography}
\end{document}